\documentclass[aps, pra, superscriptaddress , amsmath, amssymb, twocolumn, nofootinbib]{revtex4-1}

\usepackage{amsmath,amsfonts,bbm, graphicx, color}
\usepackage{perpage}
\MakePerPage{footnote}
\usepackage{amssymb}
\usepackage{appendix}
\usepackage[colorlinks]{hyperref}
\usepackage[figure,table]{hypcap}
\usepackage[textsize=tiny]{todonotes}
\usepackage{mathtools}

\hypersetup{
	bookmarksnumbered,
	pdfstartview={FitH},
	citecolor={green},
	linkcolor={darkblue},
	urlcolor={darkblue},
	pdfpagemode={UseOutlines}}
\definecolor{darkgreen}{RGB}{20,100,20}
\definecolor{darkblue}{RGB}{0,0,130}
\definecolor{darkred}{rgb}{.8,0,0}

\def\>{\rangle}
\def\<{\langle}

\def\>{\rangle}
\def\<{\langle}

\def \be{\begin{equation}}
\def \ee{\end{equation}}
\def \beq{\begin{equation}}
\def \eeq{\end{equation}}
\def \bea{\begin{eqnarray}}
\def \eea{\end{eqnarray}}

\renewcommand{\[}{\begin{equation}}
\renewcommand{\]}{\end{equation}}

\newcommand{\ket}[1]{|#1\rangle}
\newcommand{\bra}[1]{\langle#1|}

\newcommand{\abs}[1]{|#1|}

\definecolor{darkred}{rgb}{.8,0,0}
\definecolor{magenta}{RGB}{255,0,255}
\definecolor{green}{rgb}{.2,.6,.2}

\usepackage{soul}
\newcommand{\nn}{\nonumber}

\begin{document}

\title{Communication between inertial observers with partially correlated reference frames}

\author{Mehdi~Ahmadi}
\email{mehdi.ahmadi@fuw.edu.pl}
\affiliation{Institute of Theoretical Physics, University of Warsaw, Pasteura 5, 02-093 Warsaw, Poland}
\author{Alexander~R.~H.~Smith}
\email{a14smith@uwaterloo.ca}
\affiliation{Department of Physics \& Astronomy, University of Waterloo, Waterloo, Ontario Canada N2L 3G1}
\affiliation{Department of Physics \& Astronomy, Macquarie University, Sydney NSW 2109, Australia}
\author{Andrzej~Dragan}
\email{dragan@fuw.edu.pl}
\affiliation{Institute of Theoretical Physics, University of Warsaw, Pasteura 5, 02-093 Warsaw, Poland}

\date{\today}

\begin{abstract}
In quantum communication protocols the existence of a shared reference frame between two spatially separated parties is normally presumed. However, in many practical situations we are faced with the problem of misaligned reference frames. In this paper, we study communication between two inertial observers who have partial knowledge about the Lorentz transformation that relates their frames of reference. Since every Lorentz transformation can be decomposed into a pure boost followed by a rotation, we begin by analysing the effects on communication when the parties have partial knowledge about the transformation relating their frames, when the transformation is either a rotation or pure boost. This then enables us to investigate how the efficiency of communication is affected due to partially correlated inertial reference frames related by an arbitrary Lorentz transformation. Furthermore, we show how the results of previous studies where reference frames are completely uncorrelated are recovered from our results in  appropriate limits.\end{abstract}


\maketitle

\section{Introduction}
\label{Introduction}

In most quantum communication schemes it is assumed that two spatially separated parties share a common reference frame (RF), which is necessary for encoding and decoding the desired message. As an example, suppose that Alice wishes to communicate an angle $\lambda\in[0,2\pi)$ to Bob. She can encode her message $\lambda$ by preparing a quantum harmonic oscillator in a coherent state with phase $\lambda$. Then Bob will be able to decode $\lambda$ only if he has access to the phase RF with respect to which the coherent state has been prepared. 

After relaxing the assumption of having access to a shared RF, we can proceed in two ways. On one hand, available resources can be devoted to align local RFs of the involved parties. However, despite the considerable amount of progress in the development of protocols, such as clock synchronisation and Cartesian frame alignment~\cite{SSR-RF}, maintaining aligned RFs is still a major caveat of these schemes. On the other hand, the problem of quantum communication when local RFs are not aligned is known to be equal to the problem of quantum communication through a noisy channel~\cite{SSR-RF}. One of the main goals of the recently developed \textsl{resource theory of quantum reference frames}~\cite{RTA1,RTA2,RTA3}, also known as the \textsl{quantum resource theory of asymmetry}~\cite{RTA5,RTA6,RTA} has been to develop strategies to circumvent the noise due to the misalignment of RFs without the need to establish a shared RF~\cite{QCQRF}. However, the literature is mostly concerned with the special case wherein the local RFs of involved parties are completely uncorrelated. {In this paper we consider the scenario in which Alice encodes $\lambda$ in the spin degree of freedom of a massive particle. We analyse the effect of partially correlated RFs when Bob has partial knowledge about the Lorentz transformation between his frame and Alice's.} 

Suppose Bob's RF is related to Alice's RF via a Lorentz transformation. It is known, that even if Bob has perfect information about the relation between his and Alice's  RF, the information encoded in the spin degree of freedom of a massive particle is degraded due to the generation of entanglement between spin and momentum degrees of freedom in Bob's frame \cite{PeresTerno, SP, GinAda}. In this paper we analyse the total decoherence caused by both Bob's lack of knowledge about the relation between RFs and the decoherence  due to the entangling nature of quantum Lorentz transformations \cite{Weinberg}. 

This problem has been studied  previously, however, the analysis in \cite{SteveDanny} is limited to the worst case scenario in which Bob has no information whatsoever about the relative orientation of his local RF with respect to Alice's, which makes communication using a single spin-$1/2$ particle infeasible. This is simply due to the fact that the noisy channel is a depolarizing channel that completely decoheres the state. In this paper we extend the analysis to the case wherein Alice and Bob have partial information about their local reference frames. In the presence of such partially correlated RFs Bob can access the coherence in the state that has been prepared by Alice in order to decode the information encoded by Alice in a single spin-$1/2$ particle. 

This paper is structured as follows: In section \ref{AFC} we explain how Bob's lack of knowledge can be interpreted as an extra noise over the quantum channel between Alice and Bob. Then we briefly summarise quantum Lorentz transformations for a massive particle in section \ref{QLT}. We begin section \ref{AFCLT} by analysing how the state of a quantum system is perceived by Bob who has partial knowledge about the Lorentz transformation between his and Alice's RF. In section \ref{GTRsec} we restrict ourselves to the case when the transformation between Bob's and Alice's RFs is a rotation and quantify how well Bob is able to decode a classical message sent by Alice in terms of his knowledge about the rotaion. We carry out a similar analysis in section \ref{GTBsec}, when instead the transformation between Alice's and Bob's RFs is a pure boost, taking into account the finite size of the momentum wave packet of the particle sent by Alice. These analysis enable us to study the effects of partial knowledge about an arbitrary Lorentz transformation on the efficiency of communication between Alice and Bob. In section \ref{Outsec} we discuss our results and outline possible future directions for our research. 

Throughout this paper we choose natural units, i.e. $c=\hbar=1$.


\section{Preliminaries}
\subsection{Noisy quantum channel due to misalignment of reference frames}
\label{AFC}

In this section, we summarise how lack of knowledge about the transformation between local RFs of two spatially separated parties can be treated as an additional noise on the channel between them \cite{SSR-RF}.

Consider that Alice's and Bob's local RFs are related via a unitary transformation $U(g)$, where $g$ is an element of the group $G$ formed by all possible orientations of their RFs. Suppose Alice prepares a state $\rho$ with respect to her local RF, which she then sends to Bob via a perfect quantum channel. If Bob knows the relation between his RF and Alice's, i.e. if he knows $g$, he can exactly recover the state by passively transforming it as $U(g)\rho U^{\dag}(g)$. However, if he has partial knowledge about the transformation, then the state that he perceives is a weighted averaging over the group elements $g$
\begin{align}
{\cal{G}}[\rho]=\int \mathrm{d}g\, p(g) U(g)\rho U^{\dag}(g), \label{WeightedTwirl}
\end{align}
where $p(g)$ is a probability distribution characterizing Bob's knowledge about the relative orientation of his RF with respect to Alice's. We will refer to Eq.~\eqref{WeightedTwirl} as a weighted G-twirl.
 
The quantum channel $\cal{G}$ induces noise on a perfect quantum channel between Alice and Bob. The amount of induced noise depends on how peaked the probability distribution $p(g)$ is or, in other words, how ignorant Bob is about the relation between his RF and Alice's. 

Various operational measures have been introduced in order to quantify Bob's ability in decoding a message when reference frames are misaligned. In this paper we use the quantum Fisher information (QFI) as a measure of how well Bob can distinguish between the classical messages $\lambda$ and $\lambda+\epsilon$, which Alice has encoded in a quantum system via the unitary encoding  $|\psi_{\lambda}\>=e^{-iK\lambda}|\psi\>$, where $K$ is the generator of Alice's encoding. QFI provides the upper limit on the amount information that can be extracted by Bob about the encoded parameter $\lambda$ for any given measurement, which is known as the quantum Cram\'{e}r-Rao bound~\cite{paris}. To compute Bob's QFI, we use the relation between
QFI and the Uhlmann fidelity~$\mathcal{F}$ of the two  generally mixed states ${\cal{G}}[\rho_{\lambda}]$ and ${\cal{G}[\rho_{\lambda+\epsilon}}]$, given by~\cite{BraCaves}
 \begin{align}
 F(\lambda,{\cal{G}}[\rho_{\lambda}])=\frac{8\big(1-\sqrt{\mathcal{F}({\cal{G}}[\rho_{\lambda}],{\cal{G}}[\rho_{\lambda+\epsilon}])}\big)}{\epsilon^{2}}, \label{quantumfishinfo}
\end{align}
where $\mathcal{F}(\rho_1,\rho_2)=\left[\text{Tr}(\sqrt{\sqrt{\rho_1}\rho_2\sqrt{\rho_1}})\right]^{2}\,$.  

To quantify the amount of information lost due to Bob's lack of knowledge about the transformation between his frame and Alice's, we compare $F(\lambda,{\cal{G}}[\rho_{\lambda}])$ to Bob's QFI when he has complete information about the relation between the local RFs. Under such ideal conditions Alice's unitary encoding remains intact. In this situation, for a pure initial state $|\psi\>$ and unitary encoding $U_{\lambda}=e^{-i\hat{K}\lambda}$, the Eq.~\eqref{quantumfishinfo} reduces to ~\cite{paris}
\begin{align}\label{unitaryQFI}
F(\lambda,|\psi\>)=4(\<\hat{K}^2\>-\<\hat{K}\>^2),
\end{align}
where $\<\hat{X}\>=\<\psi|\hat{X}|\psi\>$.

\subsection{Quantum Lorentz transformations}
\label{QLT}

In this section we briefly review how the state of a massive quantum particle is transformed under a Lorentz transformation. We refer the readers to \cite{Weinberg} for further details.

Consider the momentum eigenstates $|\mathbf{p},m\>$ of a particle in Alice's RF with four momentum $p=(p^0,\mathbf{p})$ and the $z$-component of spin $m$. If Bob's frame is related to Alice's via a Lorentz transformation $\Lambda$, then the momentum eigenstates transform via the so called quantum Lorentz transformation as 
\begin{align}\label{QLT1}
U(\Lambda)|\mathbf{p},m\>=\sum_{m'} D^{(j)}_{{m'm}}[\Theta_{W}(\Lambda,\mathbf{p})]|\Lambda \mathbf{p},m\>,
\end{align}
where $\Theta_{W}(\Lambda,\mathbf{p})=L^{-1}(\Lambda \mathbf{p})\Lambda L(\mathbf{p})$ is the Wigner angle associated with a Lorentz transformation $\Lambda$ acting on eigenstate of momentum $\mathbf{p}$, $L(\mathbf{p})$ is a pure boost, and $\mathbf{v}$ is the boost velocity associated with the Lorentz transformation $\Lambda$. Note that the Wigner rotation is on the spin level and does not affect the momentum degree of freedom of the particle.  In other words, we have $D^{j}_{m'm}(\theta)=\<m'|U_{\hat{\mathbf{n}}}(\theta)|m\>$, where $U_{\hat{\mathbf{n}}}(\theta)=e^{-i\theta\hat{\mathbf{n}} \cdot\mathbf{J}}$ is a spin-$j$ unitary representation of the Wigner rotation around the axis $\hat{\mathbf{n}}$ parallel to $\mathbf{v}\times\mathbf{p}$.

Since the momentum eigenstates ${|\mathbf{p},m\>}$ form an orthonormal basis for the Hilbert space of a massive particle, we can expand any state in Alice's frame as   
\begin{align}\label{psi}
\left|\Psi_A\right\rangle=\sum_m\int \mathrm{d}\mu(\mathbf{p}) \,  \psi_m(\mathbf{p})|\mathbf{p},m\>,
\end{align}
where $\mathrm{d}\mu(\mathbf{p})=(2\pi)^{-3} (2E_p)^{-1}\mathrm{d}^3\mathbf{p}$ and $E_p = \sqrt{m^2 + p^2}$. Then using Eq.~\eqref{QLT1}, the transformed state of the particle in Bob's frame is given by 
\begin{align}\label{QLT2}
U(\Lambda) \left|\Psi_A\right\rangle =\sum_{m,m'}\int \mathrm{d}\mu(\mathbf{p}) \, \psi_m\left(\mathbf{p}\right) D^{(j)}_{\small{m'm}}[\Theta_{W}]|\Lambda \mathbf{p},m\>,
\end{align} 
where we have dropped the dependence of the Wigner angle on $\Lambda$ and $\mathbf{p}$, that is $\Theta_{W}=\Theta_{W}(\Lambda,\mathbf{p})$. 

In what follows, we will assume that Alice prepares a state with respect to her RF in which the spin and momentum degrees of freedom are unentangled, that is $\psi_m(\mathbf{p})=\alpha_m \psi(\mathbf{p})$, where $\alpha_{m}$ are the probability amplitudes of the spin degree of freedom, $\psi(\mathbf{p})$ represents the momentum wave function of the particle {and both spin and momentum parts of the wave function are normalised, i.e. $\sum_{m}|\alpha_m|^2=1$ and $\int \mathrm{d}\mu(\mathbf{p}) \, |\psi_m\left(\mathbf{p}\right)|^2=1$}. Since the amount and direction of the Wigner rotation depends on the momentum of the particle, these two degrees of freedom get entangled in Bob's frame. As a consequence, even when Bob is completely aware of the Lorentz transformation between his RF and Alice's, the reduced spin state with respect to Bob's frame will be decohered \cite{PeresTerno}.

\section{Communication in the presence of partial knowledge about the Lorentz transformation}
\label{AFCLT}

\begin{figure}[t]
\includegraphics[width=\linewidth]{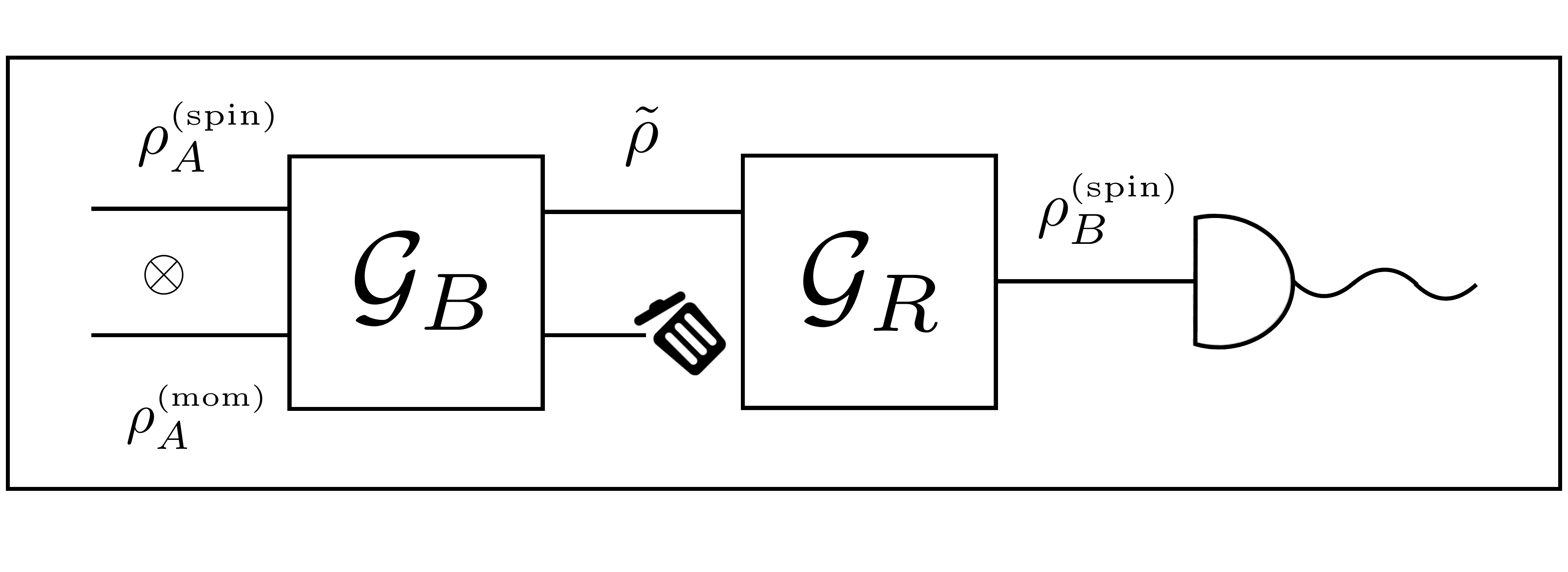}
\caption{Communication between two inertial observers in the absence of shared RFs: Alice prepares a massive particle in the product state $\rho^{\mbox{{\tiny (spin)}}}_A \otimes \rho^{\mbox{{\tiny (mom)}}}_A$, which she sends to Bob who perceives the spin part of the state as $\rho^{\mbox{{\tiny (spin)}}}_B$. The noisy quantum channels $\cal{G}_B$ and $\cal{G}_R$ represent Bob's lack of information about the relative boost and relative rotation between his local frame and Alice's. }\label{Bigpic}
\end{figure}

In the previous section, we reviewed how lack of knowledge about the relation between local RFs can be treated as a noisy channel and  how a state of a massive particle is transformed with respect to different Lorentz frames of reference. In this section we will show how the total amount of noise, resulting from the involved parties having partial information about the Lorentz transformation between their local frames, affects Bob's optimal performance in decoding a message sent by Alice; see Fig. \ref{Bigpic}.

Suppose Alice prepares a massive spin-1/2 particle in the state
\begin{align}
\ket{\Psi_A} = \sum_{m} \int \mathrm{d}\mu (\mathbf{p}) \, \alpha_m \psi(\mathbf{p})  \ket{\mathbf{p},m},
\end{align}
with $m \in \left\{-1/2,1/2\right\}$, which contains no entanglement between the spin and momentum degrees of freedom. Bob will describe the state prepared by Alice $\rho_A = \ket{\Psi_A} \bra{\Psi_A}$, with respect to his RF as
\begin{align}
\rho_B = \int \mathrm{d} \Lambda \, p\left(\Lambda\right)U\left(\Lambda\right) \rho_A U^{\dagger}\left(\Lambda\right), \label{BobState}
\end{align}
where $p\left(\Lambda\right)$ is a probability distribution satisfying $\int \mathrm{d} \Lambda \, p(\Lambda) = 1$, charcterising his knowledge about the Lorentz transformation relating his frame to Alice's.

Let us introduce $\rho_A^{\mbox{{\tiny (spin)}}}= \sum_{m,m'}\alpha_m \alpha_{m'}^*  \ket{m}\bra{m'}$ to denote the state of the spin degrees of freedom, and $\rho_A^{\mbox{{\tiny (mom)}}}= \int \mathrm{d}\mu (\mathbf{p}) \mathrm{d}\mu (\mathbf{p}') \, \psi(\mathbf{p}) \psi(\mathbf{p}')^*  \ket{\mathbf{p}}\bra{\mathbf{p}'}$ to denote the state of the momentum degrees of freedom, with respect to Alice's RF. 

The state of the spin degree's of freedom with respect to Bob RF $\rho_B^{\mbox{{\tiny (spin)}}}$, is given by a trace over the momentum degree of freedom of the state $\rho_B$ in Eq.~\eqref{BobState}
\begin{align}
 \rho^{\mbox{{\tiny (spin)}}}_{B} \!
&=  \int \mathrm{d}  \mu (\mathbf{ p}) \,  \left\langle \mathbf{p} \right|  \rho_B  \left| \mathbf{p} \right\rangle   \nonumber\\
&=  \int \mathrm{d} \Lambda \,  p(\Lambda) \int \mathrm{d}  \mu (\mathbf{p}) \,  \abs{\psi (\mathbf{p})}^2   U(\boldsymbol{\theta}_{W})\rho^{\mbox{{\tiny (spin)}}}_{A} U^{\dag}(\boldsymbol{\theta}_W), \nonumber\\\label{BobState2}
\end{align}
where $U(\boldsymbol{\theta}_W )=e^{i \boldsymbol{\theta}_W(\Lambda, \mathbf{p}) \cdot \mathbf{J} }$ and $\mathbf{J}=\frac{1}{2}\boldsymbol{\sigma}=
\frac{1}{2}(\sigma_x,\sigma_y,\sigma_z)$, where $\sigma_i$'s are the Pauli matrices.

Any Lorentz transformation can be decomposed into a pure boost followed by a rotation: $\Lambda = R(\boldsymbol{\psi}) L(\mathbf{v})$, where $\hat{\boldsymbol{\psi}}$ and $\psi = \left\|\boldsymbol{\psi}\right\| \in [0,\pi)$ describe the direction and amount of rotation respectively, and $\mathbf{v}$ is the boost velocity associated with the pure boost $L(\mathbf{v})$. This allows us to express the Wigner rotation as
\begin{align}
e^{i \boldsymbol{\theta}_W\left( \Lambda, \mathbf{p} \right) \cdot \mathbf{J}} = e^{ i \boldsymbol{\psi} \cdot \mathbf{J} } e^{ i  \boldsymbol{\phi}(\mathbf{v}, \mathbf{p}) \cdot \mathbf{J}},
\end{align}
where $\boldsymbol{\phi}(\mathbf{v},\mathbf{p})$ is a vector characterizing the Wigner rotation associated with $L(\mathbf{v})$ acting on a particle of momentum $\mathbf{p}$. 

We may express the measure appearing in Eq.~\eqref{BobState2} as $p(\Lambda) \mathrm{d} \Lambda = p(\boldsymbol{\psi},\mathbf{v}) \mathrm{d} \boldsymbol{\psi} \mathrm{d}\mathbf{v}$. This decomposition suggests we should factor Bob's knowledge about the Lorentz transformation as $p(\boldsymbol{\psi},\mathbf{v}) = g(\boldsymbol{\psi})h(\mathbf{v})$, where $g(\boldsymbol{\psi})$ and $h(\mathbf{v})$ are distributions that characterize Bob's knowledge about the rotation $\boldsymbol{\psi}$ and the boost velocity $\mathbf{v}$ respectively. With this, Eq.~\eqref{BobState2} can be written as
\begin{align}
\rho^{\mbox{{\tiny (spin)}}}_B =   \int \mathrm{d}\boldsymbol{\psi} \, g(\boldsymbol{\psi}) e^{ i \boldsymbol{\psi} \cdot \mathbf{J} } \tilde{\rho} e^{- i \boldsymbol{\psi} \cdot \mathbf{J} },
\label{BobState3}
\end{align}
where 
\begin{align}\label{GTBM}
\tilde{ \rho}=\int \mathrm{d} \mathbf{v}  \, h(\mathbf{v}) \int\mathrm{d}\mu ( \mathbf{p}) \,  |\psi(\mathbf{p})|^2  e^{i \boldsymbol{\phi}(\mathbf{v},\mathbf{p}) \cdot \mathbf{J} } \rho^{\mbox{{\tiny (spin)}}}_{A} e^{- i  \boldsymbol{\phi}(\mathbf{v},\mathbf{p}) \cdot \mathbf{J} },
\end{align}
is the weighted G-twirled state when the Lorentz transformation is a pure boost $\Lambda = L(\mathbf{v})$; see Fig.~\ref{Bigpic}. 

In what follows, we will compute the weighted G-twirl over pure rotations and pure boosts separately. In both cases we compute the QFI to quantify Bob's ability in decoding $\lambda$.


\begin{figure}[t]
\centering
\includegraphics[width=0.45\textwidth]{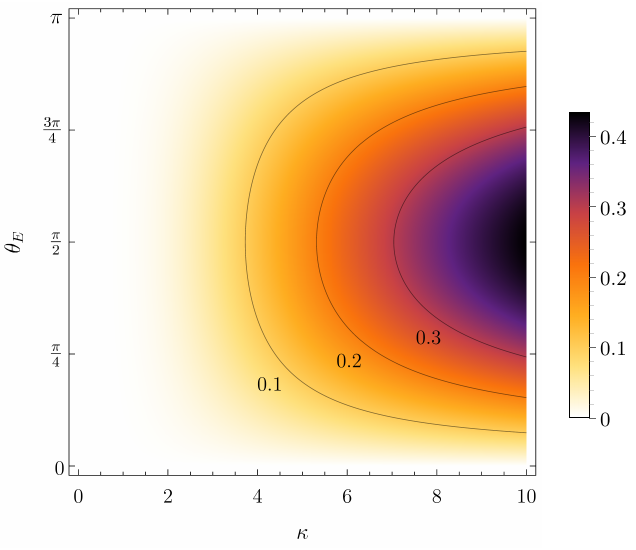}
\caption{Bob's QFI determined in Eq.~\eqref{QFIrotaions} as a function of his lack of information $\kappa$ about the rotation relating his RF to Alice's and Alice's choice of encoding direction $\theta_E$ quantifying how well he can measure the parameter $\lambda$. We observer the QFI is peaked around $\pi/2$ and vanishes as $\kappa$ goes to zero, which corresponds to the limit when Bob's has no knowledge of the relation of his RF to Alice's.}
\label{fig:RotaitonPlots}
\end{figure}
\subsection{Weighted G-twirling over rotations}\label{GTRsec}
In this section, we analyse the weighted G-twirl of the state $\rho^{\mbox{{\tiny (spin)}}}_{A}$ over the group of rotations $SO(3)$. We label each element of $SO(3)$ by the axis-angle pair $(\hat{\mathbf{n}},\psi)$, where $\hat{\mathbf{n}}$ is a unit vector indicating the direction of the axis of rotation and $\psi \in [0,\pi)$ is an angle describing the amount of rotation. With such a parametrization, the unitary representation of a group element is
\be
e^{i\psi  \hat{\mathbf{n}} \cdot \mathbf{J}}=\cos\left(\frac{\psi}{2}\right)\mathbb{I}+2 i\sin\left(\frac{\psi}{2}\right)\hat{\mathbf{n}}\cdot\mathbf{J}.
\ee

As the group $SO(3)$ is diffeomorphic to {the real projective space} $\mathbb{R} \mathbf{P}^3$, we may alternatively identify elements of $SO(3)$ with points on a 3-sphere with antipodal points identified\footnote{Consider a solid ball in $\mathbb{R}^3$ of radius $\pi$. Each point in the ball corresponds to a rotation around the axis defined by the point and the origin, by a rotation angle equal to the distance between the point and the origin. Rotations on opposite sides of the surface of the ball represent the same rotation. Thus we identify antipodal points of the ball, which results in the real projective space $\mathbb{R} \mathbf{P}^3$.}. The advantage of this will be that we can characterize Bob's knowledge about the relation of his RF to Alice's as a von Mises-Fisher distribution, {which is a natural generalization of a Gaussian distribution to a sphere; see appendix \ref{2} for details}. The 3-sphere can be defined as $S^3 = \{ \mathbf{x} \in  \mathbb{R}^4 \ | \   \mathbf{x} \cdot \mathbf{x} =1 \}.$ To connect points on $S^3$ with elements of $SO(3)$ we introduce a hyperspherical coordinate system $(\psi,\theta, \phi)$ with $\psi, \theta \in [0,\pi)$ and $\phi\in [0,2 \pi)$, related to the usual Cartesian coordinates by
\begin{align}
\mathbf{x} = 
\begin{pmatrix}
\cos \psi \\
 \sin \psi \cos \theta \\
  \sin \psi \sin \theta \cos \phi \\
  \sin \psi \sin \theta \sin \phi
\end{pmatrix}.
\end{align}
The point $\boldsymbol{\psi} = (\psi,\theta,\phi)$ represents a rotation around the axis $\hat{\mathbf{r}} = (\sin \theta \cos \phi, \sin \theta \sin \phi,\cos \theta)^T$ through an angle $\psi$.

We choose to characterize Bob's knowledge about the rotation that takes his frame to Alice's, by the von Mises-Fisher distribution on $S^3$. Without loss of generality, we choose the distribution to be centred around the identity rotation labeled by the point $\boldsymbol{\psi}_0 =(0,0,0)$, as if it was centred around any other point Bob would be able to rotate his RF such that this was the case. With this choice, the probability density function in hyperspherical coordinates takes the form
\begin{align}
g\left(\psi,\theta,\phi\right) = \frac{\kappa}{4 \pi^2 I_1 \! \left(\kappa \right)}  \exp \left[ \kappa \cos \psi \right], \label{RotaionDistribution}
\end{align}
where $I_1(\kappa)$ denotes the first-order modified Bessel function of the first kind and $\kappa>0$ is known as the concentration of the distribution; as $\kappa$ increases the distribution becomes more peaked around the rotation $\boldsymbol{\psi}_0$, and in the limit $\kappa \rightarrow 0$ the distribution limits to the uniform distribution on $SO(3)$.

Using the von Mises-Fisher distribution to define a weighted G-twirl over the group of rotations, we find
\begin{align}
\mathcal{G}_{R} \left[\rho^{\mbox{{\tiny (spin)}}}_{A}\right] &= \int \mathrm{d} \boldsymbol{\psi} \, g\left(\psi,\theta,\phi\right) e^{i\psi  \hat{\mathbf{n}} \cdot \mathbf{J}} \rho^{\mbox{{\tiny (spin)}}}_{A}e^{-i\psi  \hat{\mathbf{n}} \cdot \mathbf{J}} \nn \\
&= \left(1 -\frac{3G(\kappa)}{\kappa} \right) \rho^{\mbox{{\tiny (spin)}}}_{A} + \frac{G\left(\kappa\right)}{ \kappa} \sum_{i} \sigma_j\rho_A^{\mbox{{\tiny (spin)}}}\sigma_j, \label{RotaionTwirledState}
\end{align}
where $\mathrm{d} \boldsymbol{\psi} =\sin^2 \psi \sin \theta \mathrm{d} \psi \mathrm{d} \theta \mathrm{d} \phi$ and $G\left(\kappa\right) = I_2(\kappa)/I_1(\kappa)$ is the population mean resultant length of the von Mises-Fisher distribution on $S^3$.
We note that the quantum channel $\mathcal{G}_{R}$ in \eqref{RotaionTwirledState}
  is a \textsl{depolarising channel}.  In such a channel either with probability $1-p$ the qubit remains intact or one of the three types of errors, i.e. bit flip error, phase flip error or both, with equal  probability $p=3G(\kappa)/\kappa$ occurs. 
 
We suppose Alice encodes the real number $\lambda$ via the unitary encoding $|\psi_{\lambda}\>=e^{-i\lambda\hat{\mathbf{E}} \cdot \mathbf{J}}|0\>$, where $J_z|0\>=\frac{1}{2}|0\>$ and $\hat{\mathbf{E}} = (\sin \theta_E \cos \phi_E, \sin \theta_E \sin \phi_E, \cos \theta_E)^T$ is a unit vector representing Alice's choice of encoding, so that $\rho^{\mbox{{\tiny (spin)}}}_{A} = |\psi_{\lambda}\> \<\psi_{\lambda}|$. {Note that $\theta_E$ is the angle between Alice's encoding direction and the direction along which Bob's knowledge is concentrated.}  We compute the QFI of the state $\mathcal{G}_{R} [\rho^{\mbox{{\tiny (spin)}}}_{A}]$, using Eqs. \eqref{quantumfishinfo} and \eqref{RotaionTwirledState}, as a measure of how well Bob is able to determine $\lambda$, with the result
\begin{align}
F(\lambda,{\cal{G}}_{\mbox{\tiny{R}}}[\rho_{\lambda}]) = \sin^2 \theta_E \left(1-\frac{4G(\kappa)}{\kappa}\right)^2. \label{QFIrotaions}
\end{align} 

{From Eq.~\eqref{QFIrotaions}, we find that QFI is maximised when $\theta_E=\frac{\pi}{2}$, which means that the optimal encoding direction is orthogonal to the direction the distribution characterizing Bob's knowledge is peaked in.} In addition we observe, as expected, in the limit when Bob has no knowledge about the rotation between his frame and Alice's $\kappa \to 0$, the QFI vanishes $F(\lambda,{\cal{G}}_{\mbox{\tiny{R}}}[\rho_{\lambda}]) \to 0$; and in the limit when he knows the rotation exactly $\kappa \to \infty$, the QFI limits to $F(\lambda,{\cal{G}}_{\mbox{\tiny{R}}}[\rho_{\lambda}]) \to \sin^2 \theta_E$.

Our results in the latter limit completely agree with the results obtained in \cite{QPEIRF}. We also note that, in the context of noisy quantum metrology, similar results have been observed. Specifically, the authors of \cite{Acin} showed that the precision of quantum parameter estimation can be improved significantly when the noise is concentrated along a direction perpendicular to the plane in which the system is evolving.



\subsection{Weighted G-twirling over pure boosts}\label{GTBsec}

Our aim in this section is to compute a weighted G-twirling over pure boosts, that is to evaluate the integration in Eq.~\eqref{GTBM} in which $\tilde{\rho}$ is defined. 
Making use of the identity  $e^{i a(\hat{\mathbf{n}} \cdot \boldsymbol{\sigma})} = \mathbb{I} \cos{a}  + i (\hat{\mathbf{n}} \cdot \boldsymbol{\sigma}) \sin{a}$, we may express the Wigner rotation corresponding to a pure boost $L(\mathbf{v}$) on a particle of momentum $\mathbf{p}$ as
\begin{align}\label{WeignerExpansion}
e^{i \boldsymbol{\phi} \cdot \mathbf{J}} = \mathbb{I} \cos \frac{\phi}{2} + 2 i \left(\hat{\boldsymbol{\phi}} \cdot \mathbf{J}\right) \sin \frac{\phi}{2},
\end{align}
where we have suppressed the dependence of $\boldsymbol{\phi}(\mathbf{v},\mathbf{p})$ on $\mathbf{v}$ and $\mathbf{p}$, so that $\boldsymbol{\phi} = \boldsymbol{\phi}(\mathbf{v}, \mathbf{p})$; the amount of rotation $\phi$ and the axis of rotation $\hat{\boldsymbol{\phi}}$ are given in appendix \ref{1}. 
Substitution of Eq.~\eqref{WeignerExpansion} into Eq.~\eqref{GTBM} yields
\begin{align}
&\tilde{\rho}=\int \mathrm{d} \mathbf{v}  \, h(\mathbf{v}) \int\mathrm{d}\mu \, ( \mathbf{p})\left|\psi(\mathbf{p})\right|^2  \Bigg[ \cos^2 \frac{\phi}{2} \rho_A^{({\rm spin})} \nonumber \\
& +i \sin \phi  \left[\hat{\boldsymbol{\phi}}\cdot \mathbf{J}, \rho_A^{({\rm spin})}\right] + 4 \sin^2\frac{\phi}{2}   \left(\hat{\boldsymbol{\phi}}\cdot \mathbf{J} \right)\rho_A^{({\rm spin})} \left( \hat{\boldsymbol{\phi}}\cdot \mathbf{J}\right) \Bigg]. \label{integrand1}
\end{align} 

We assume that the particle Alice is using to communicate with Bob is approximately at rest in her RF, which amounts to assuming $p/m \ll 1$. This enables us to expand $\cos^2\phi$ and $\sin^2 \phi$ appearing in Eq.~\eqref{integrand1} to second order in $p/m$ as
\begin{align}
\cos \phi &\approx 1 - \frac{1}{2} F\left(v\right)^2 \left(\frac{p}{m}\right)^2 \left(1-\left(\hat{\mathbf{v}} \cdot \hat{\mathbf{p}}\right)^2  \right), \label{cosphi}\\
\sin \phi \, \hat{\boldsymbol{\phi}} &\approx \left[F\left(v\right) \left(\frac{p}{m}\right)  - \frac{1}{2} F\left(v\right)^2 \left(\frac{p}{m}\right)^2 \, \hat{\mathbf{v}} \cdot \hat{\mathbf{p}}  \right] \hat{\mathbf{v}}\times\hat{\mathbf{p}}, \label{sinphi}
\end{align}
where we have defined $F(v):= v/\left(1+\sqrt{1+v^2}\right)$. 

We choose $h(\mathbf{v})$, the probability distribution characterising Bob's knowledge about the boost velocity relating his RF to Alice's, to be of the form $h(\mathbf{v}) = h_1(\theta_v,\phi_v)  h_2(v)$, where $ h_1(\theta_v,\phi_v)$ is a distribution over azimuthal and polar angles $\theta_v$ and $\phi_v$ indicating the direction of the boost with respect to Bob's RF and $h_2(v)$ is a distribution over the magnitude of the boost velocity $v = \left\| \mathbf{v}\right\|$. We choose $h_1(\theta_v,\phi_v)$ to be a von Mises-Fisher distribution on $S^2$ centred around Bob's $z$-axis $\hat{\mathbf{v}}_0=(0,0,1)^T$ and  $h_2(v)$ to be a bump function on the interval $[0,1)$:
\begin{align}
h_1(\theta_v,\phi_v)&=\frac{1}{(2\pi)^{\frac{3}{2}}}\left(\frac{\kappa_v}{2}\right)^{\frac{1}{2}}e^{\kappa_v \, \hat{\mathbf{v}}_0 \cdot \hat{\mathbf{v}}}, \\
h_2(v) &= \frac{1}{N(\Delta)} \exp\left[-\frac{1}{\Delta^2 (1-v^2)}\right],
\end{align}
where $\hat{\mathbf{v}} = (\sin \theta_v \cos \phi_v, \sin \theta_v \sin \phi_v, \cos \theta_v)^T$ and $N(\Delta) := \int_0^1dv \, \exp[-1/(\Delta^2 (1-v^2))]$. The parameters $\kappa_\nu$  and $\Delta$ determine how well Bob knows the direction and magnitude of the boost velocity relating his RF to Alice's.

Similarly, we suppose Alice prepares the momentum wave packet $\psi(\mathbf{p})$ such that the probability distribution characterizing the particle's momentum is of the form $|\psi(\mathbf{p})|^2= f_1(\theta_p,\phi_p)  f_2(p)$, where $\theta_p$ and $\phi_p$ are again azimuthal and polar angular respectively indicating the direction of particle's momentum as prepared by Alice and $p = \left\| \mathbf{p}\right\|$ is the magnitude of momentum. We choose $f_1(\theta_p,\phi_p)$ to be a von Mises-Fisher distribution on $S^2$ centred around  $\hat{\mathbf{p}}_0=(1,0,0)^T$ and  $f_2(p)$ to be sufficiently peaked around momentum $p_0$ that we may take it to be an appropriately normalized delta function:
\begin{align}
f_1(\theta_p,\phi_p) &=\frac{1}{(2\pi)^{\frac{3}{2}}}\left(\frac{\kappa_p}{2}\right)^{\frac{1}{2}}e^{\kappa_p \, \hat{\mathbf{p}}_0 \cdot \hat{\mathbf{p}}}, \\
f_2(p) &=(2\pi)^{3}(2E_{p})\delta(p-p_0), \label{boostDistribution}
\end{align}
where $\hat{\mathbf{p}} = (\sin \theta_p \cos \phi_p, \sin \theta_p \sin \phi_p, \cos \theta_p)^T$ and the parameter $\kappa_p$ determines how concentrated the distribution is around the direction $\hat{\mathbf{p}}$.

We begin by integrating Eq.~\eqref{integrand1} over $\theta_p$, $\phi_p$, $\theta_v$ and $\phi_{p}$, using the expansions in Eqs.~\eqref{cosphi} and \eqref{sinphi}, and our choice of $h(\mathbf{v})$ and $\left|\psi(\mathbf{p})\right|^2$ above; the integration in Eq.~\eqref{integrand1} results in
\begin{align}
\tilde{\rho}=& \, c_1\rho_A^{\mbox{{\tiny (spin)}}}+i c_2 [\sigma_y,\rho_A^{\mbox{{\tiny (spin)}}}]+\sum_{j=1}^{3} C_{j}\sigma_{j}\rho_A^{\mbox{{\tiny (spin)}}}\sigma_{j}.
\label{AfterBoostTwirl}
\end{align}
The coefficients $c_1$, $c_2$ and $C_{j}$ are given in appendix \ref{Cs} in terms of $\kappa_v$, $\kappa_p$ and $T_n:=T^{(p)}_{n}T^{(v)}_{n}$, where $T^{(v)}_{n}:=\int_0^1 \mathrm{d}v \,  v^{2} (F(v))^n h_2 (v)$ and $T^{(p)}_{n}:=\left(\frac{p_0}{m}\right)^n$ for $n=1,2$.

Let us now analyse the G-twirled state \eqref{GTBM} in different limits of $\kappa_p$ and $\kappa_v$. First, when the direction of momentum distribution and Bob's knowledge about the direction of boost are highly peaked, i.e. in the limit of $(\kappa_{p},\kappa_v)\rightarrow\infty$, the state $\tilde{\rho}$ can be written as
\begin{align}\label{rho0}
\tilde{\rho}_0= 
\rho_A^{\mbox{{\tiny (spin)}}}+i \frac{T_1}{2} [\sigma_2,\rho_A^{\mbox{{\tiny (spin)}}}],
\end{align}
which is simply a rotation through through an angle of $T_1$ around $y$-axis\footnote{{Using the formula  $e^A B e^{-A}=B+[A,B]+...$, one can easily check that $e^{i\frac{\theta}{2}\sigma_2}
\rho\,
e^{-i\frac{\theta}{2}\sigma_2}=\rho+i\frac{\theta}{2}[\sigma_2,\rho]+{\cal{O}}(\theta^2)$.}}. This means that there are no decoherence effects in this limit. This is a result of the limit $\kappa_p\rightarrow\infty$ corresponding to the case in which Alice prepares the momentum degree of freedom in a momentum eigenstate; it is known that under such circumstances a Lorentz boost does not entangle the spin and momentum degrees of freedom, and consequently the noise due to this effect is not present~\cite{SteveDanny}.

Secondly, let us suppose that that Bob is completely unaware of the direction of the boost between his RF and Alice's, i.e. $\kappa_v\rightarrow 0$. Under such circumstances the state $\tilde{\rho}$ can be written as 
\begin{align}
\tilde{\rho}_1 &=  \left(1+ \frac{T_2}{6}\right) \rho_A^{\mbox{{\tiny (spin)}}}- \left(\frac{T_2}{12}\right) \ \bigg[\frac{2}{\kappa_p}H(\kappa_p)\sigma_1 \rho_A^{\mbox{{\tiny (spin)}}}\sigma_1 \nonumber\\
&\quad +\left(1-\frac{1}{\kappa_p}H(\kappa_p)\right)\left(\sigma_2 \rho_A^{\mbox{{\tiny (spin)}}}\sigma_2+\sigma_3 \rho_A^{\mbox{{\tiny (spin)}}}\sigma_3\right)\bigg] \label{rho1}, 
\end{align}
where $H(\kappa_p):=\coth \kappa_p - 1/\kappa_p$.

{If we further assume that Alice prepares the momentum wave packet in such a way that the momentum distribution is uniform in all directions, i.e. $\kappa_p\rightarrow 0$, we find}
\begin{align}
\tilde{\rho}_2= \, \left(1-\frac{T_2}{6}\right) \rho_A^{\mbox{{\tiny (spin)}}}+ \left(\frac{T_2}{18}\right)\sum_{j=1}^{3} \sigma_j \rho_A^{\mbox{{\tiny (spin)}}} \sigma_j, \label{rho2}
\end{align}
which we identify as a depolarising channel with probability $p=\frac{T_2}{6}$.

We are finally in position to compute Bob's quantum Fisher information when he has partial information about the boost transformation between his frame and Alice's. Similar to the previous section,  we suppose Alice encodes the real number $\lambda$ via a unitary encoding $|\psi_{\lambda}\>=e^{-i\lambda\hat{\mathbf{E}} \cdot \mathbf{J}}|0\>$, where $J_z|0\>=\frac{1}{2}|0\>$ and $\hat{\mathbf{E}} = (1,0,0)$, so that $\rho^{\mbox{{\tiny (spin)}}}_{A} = |\psi_{\lambda}\> \<\psi_{\lambda}|$. Then Bob's quantum Fisher information for states $\tilde{\rho}_0$, $\tilde{\rho}_1$, and $\tilde{\rho}_2$ defined above, read as 
\begin{align}\label{QFI012}
F(\lambda,\tilde{\rho}_0)&=1\nonumber\\
F(\lambda,\tilde{\rho}_1)&= \Bigg(1 - \frac{T_2}{6} \left( 1 + \frac{1}{\kappa_p}H(\kappa_p)\right)\Bigg)^2 \nonumber\\
F(\lambda,\tilde{\rho}_2)&=\left(1- 
\frac{2}{9} T_2\right)^2.
\end{align}

\begin{figure}[t]
\centering
\includegraphics[width=0.45\textwidth]{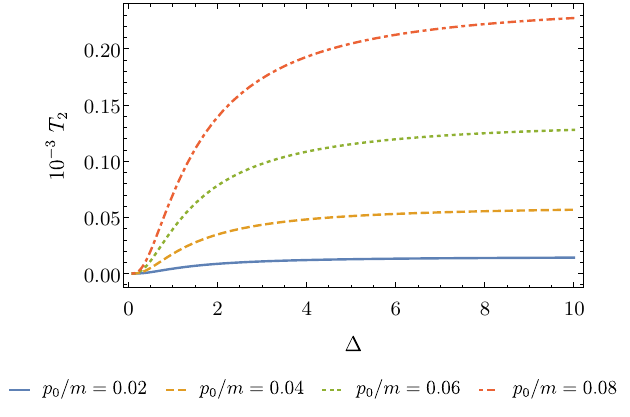}
\caption{$T_2$, which quantifies the noise in the channel between Alice and Bob is plotted as a function of Bob's knowledge $\Delta$ about the relative boost velocity between his RF and Alice's for different values of $p_0/m$. As expected, as $\Delta$ increases, corresponding to Bob becoming more uncertain about the boost velocity, $T_2$ increases. We also observe the smaller $p_0/m$ is, the  less noisy the channel is.}
\label{I2}
\end{figure}

The quantum Fisher information depends on the parameter $T_2$, which by inspection of Eqs. \eqref{rho1} and \eqref{rho2} quantifies the noise in the communication channel between Alice and Bob resulting from Bob's lack of knowledge about the orientation of his RF with respect to Alice's. Note that  in Fig. \ref{I2}, we have plotted $T_2$ in terms of Bob's knowledge about the relative boost velocity $\Delta$.  As can be seen in Fig. \ref{I2}, the reduction of quantum Fisher information due to Bob's lack of knowledge about the boost is negligible. {This can be understood by noting that the G-twirled states in Eqs. \eqref{rho1} and \eqref{rho2} up to second order in ${p_0}/{m}$ can be written as
\begin{align}
\tilde{\rho}_1&=\rho_A^{\mbox{{\tiny (spin)}}}+{\cal{O}}\left(\frac{p_0}{m}\right)^2,\nonumber\\
\tilde{\rho}_2&=\rho_A^{\mbox{{\tiny (spin)}}}+{\cal{O}}\left(\frac{p_0}{m}\right)^2.
\end{align}}
{This simply means that the G-twirling operation with respect to the relative boost between the two RFs, as given in \eqref{GTBM}, does not decohere the state prepared by Alice up to second order in $p_0/m$. Therefore, the only noise due to Bob's lack of information about the relative Lorentz transformation is due to the G-twirling with respect to rotations. In section \ref{GTRsec} we carefully analysed how Bob's ability in decoding $\lambda$ decreases due to his lack of knowledge about the relative rotation between RFs, where we also showed  how Alice can optimally encode $\lambda$ in the state of a spin-$1/2$ particle.} 

As an example, let us suppose that Alice uses an electron as her spin-$1/2$ particle with mass $m_e\approx 9.1\times 10^{-31} \,{\rm kg}$, and suppose the velocity of the electron is approximately  $0.1 \,\rm{m/s}$, so that $p_0/m \simeq 0.01 \ll 1$ and thus the approximations in Eq.~\eqref{sinphi} are still valid.

Finally, it is worth mentioning that rather than expanding $\cos^2\phi$ and $\sin^2 \phi$ appearing in Eq.~\eqref{integrand1} in $p/m$, we could have instead assumed that Bob has sufficient knowledge about the boost velocity $v \ll 1$, i.e. assuming that his knowledge is sharply peaked around $v_0=0$, which would have allowed us to expand $\cos^2\phi$ and $\sin^2 \phi$ in $v$. After repeating all the calculations in this section, we reach the same conclusion that Bob's lack of knowledge about the relative boost can be safely ignored.

\section{Discussions and outlook}
\label{Outsec}

In this paper, we analysed a communication scenario in which the involved parties have partial information about the Lorentz transformation that relates their RFs. Motivated by the fact that any Lorentz transformation can be written as a boost followed by a rotation, we investigated the effect of partial knowledge about pure rotations and pure boosts separately, while carefully taking into account the uncertainty in direction and magnitude of the momentum of the initial state. We used the QFI as an operational measure for the quality of communication between the two parties and  showed how the results of previous studies~\cite{QPEIRF} are recovered from our results when a suitable limit is approached. In particular, for the situation in which the two local RFs are related via a pure rotation, we find that the optimal encoding direction is orthogonal to the direction along which Bob's knowledge is peaked. For RFs related via a pure boost, we conclude that the effect of decoherence can be safely ignored up to second order in $p_0/m$. Here, $m$ is the mass of the transmitted particle and $p_0$ corresponds to value where the distribution of its momentum magnitude is peaked.

We emphasize, that although we have chosen specific distributions to characterise Bob's knowledge of how his RF relates to Alice's, all the results presented can be easily generalized to arbitrary distribution and expressions evaluated numerically. Our reason for choosing the distribution we did, was to obtain an analytic expression for the state prepared by Alice with respect to Bob's RF. More general distributions are not expected to exhibit qualitative features that are not present in the distributions considered.

In a forthcoming work we will investigate the possibility of Alice using the momentum degree of freedom to encode information about her RF, which Bob can use to improve his estimation of $\rho_A^{\mbox{{\tiny (spin)}}}$. 
As entanglement is produced between the momentum and spin degrees of freedom in transforming the state of a qubit from Alice's RF to Bob's RF, Alice should be able to use both degrees of freedom as the environment to encode her desired message more efficiently.

In this paper, we studied the effects of partially correlated RFs on the efficiency of communication when information is encoded in the spin  degree of freedom of a  massive particle. Firstly, it would be of practical interest to repeat our analysis for the case wherein Alice uses a photonic system to encode information. Such an analysis will generalise the previous studies of alignment-free communication~\cite{Aolita} to the case of inertial observers with partially correlated RFs. Secondly, it would be of interest to analyse the effect of partially correlated RFs in the context of the violation of Bell inequalities~\cite{VOB1, VOB2, VOB3}.

Recently, the amount of coherence in generally mixed quantum states has been operationally quantified \cite{Plenio, RTA7}. Also it has been shown that the coherence of a noisy quantum channel is related to the average change in purity averaged over input pure states \cite{Flammia}. As a future line of research, we are interested in exploiting these operational measures in order to devise optimal communication scenarios for the case of partially correlated RFs. The noisy channel caused by Bob's lack of information in such a scenario is not a completely incoherent channel, as opposed to the case wherein Bob is completely ignorant about the relation between the local RFs. In the latter case, it is known that the most coherent state is the most resourceful state for alignment-free communication \cite{RTA1,RTA3}. One of the questions we would like to answer is to find the optimal state for partially correlated RFs.

Last but not least, our analysis has interesting connections with disparate areas in foundations of quantum mechanics and quantum information theory. To name a few: the role of reference frames in quantum optical interferometry~\cite{Rafal}, the decoherence caused due to the quantum nature of reference frames such as quantum phase reference frames~\cite{Terrydeg} or directional quantum reference frames~\cite{PY,UnitaryDeg}, and conditional probability interpretation of time in quantum mechanics~\cite{PM}. \\


\emph{Acknowledgements}:\
We thank Daniel Terno for useful discussions and comments. M.~A. and A.~D. acknowledge support from the National Science Center,
Sonata BIS Grant No. 2012/07/E/ST2/01402.\\

\appendix

\section{The von Mises-Fisher distribution}
\label{2}

We summarize here the basic properties of the von Mises-Fisher distribution used in Eq. \eqref{RotaionDistribution} to characterize Bob's knowledge of his relation to Alice's reference frame and in Eq. \eqref{boostDistribution} to define the momentum distribution of the state prepared by Alice and Bob's knowledge of the boost direction relating his RF to Alice's. The von Mises-Fisher distribution, in some sense, can be thought of as the natural generalization of a normal distribution to a $(p-1)$-sphere, with the concentration $\kappa$ playing the role of the inverse of the standard deviation of the normal distribution.

A random unit vector $\mathbf{x}$ has the $(p-1)$ von Mises-Fisher distribution if it's probability density function, with respect to the uniform distribution, has the form
\begin{align}
f(\mathbf{x}) = \left( \frac{\kappa}{2} \right)^{p/2 -1} \frac{1}{\Gamma(p/2) I_{p/2-1}(\kappa)} \exp \left(\kappa \boldsymbol{\mu} \cdot \mathbf{x} \right), \label{vMFdist}
\end{align}
where $\kappa \geq 0 $, $\left\| \mu \right\| =1$, and $I_{\nu}$ denotes the modified Bessel function of the first kind and order $\nu$. As the probability density function in Eq.~\eqref{vMFdist} is symmetric around $\boldsymbol{\mu}$, the mean direction of $\mathbf{x}$ is $\boldsymbol{\mu}$. $\kappa$ is the concentration of the distribution---the greater $\kappa$ the more peaked the distribution is around the mean direction $\boldsymbol{\mu}$. 

The mean resultant length of a random unit vector  $\mathbf{x}$ distrbuted acording to Eq.~\eqref{vMFdist} is
\begin{align}
\bar{\rho} := \left(\sum_{n=1}^p \left\langle x_i\right\rangle^2 \right)^{1/2} = \frac{I_{p/2}(\kappa)}{I_{p/2 -1}(\kappa)}.
\end{align}
When $p=3$, as was the case in Eq.~\eqref{boostDistribution}, the mean resultant length has the simple form $H(\kappa) = \coth \kappa -1/\kappa$, which appears throughout the paper; specifically in appendix \ref{Cs} where we explicitly state the coefficients appearing in Eq.~ \eqref{AfterBoostTwirl}. When $p=4$ the mean resultant length is $G\left(\kappa\right) = I_2(\kappa)/I_1(\kappa)$, which was introduced just below Eq.~\eqref{RotaionTwirledState}.

More details on the von Mises-Fisher distribution can be found in~\cite{DSB}.

\section{Wigner rotation for pure boosts}
\label{1}

The Wigner rotation for a spin-$1/2$ particle with momentum $\mathbf{p}$ and mass $m$, resulting from a pure boost $L(\mathbf{v})$ is a rotation by an amount $\phi$ around the axis $\hat{\boldsymbol{\phi}}$, both of which are given by \cite{Soo,Halpern}
\begin{widetext}
\begin{align}
\cos \phi &= \frac{\sqrt{v^2+1} + \sqrt{\tilde{p}^2+1} + v \tilde{p} \left(\hat{\mathbf{v}} \cdot \hat{\mathbf{p}}\right) + \left(\sqrt{v^2+1} -1 \right)\left(\sqrt{\tilde{p}^2+1} - 1\right)\left(\hat{\mathbf{v}} \cdot \hat{\mathbf{p}}\right)^2}{1 + \sqrt{v^2+1} \sqrt{\tilde{p}^2+1} + v \tilde{p} \left(\hat{\mathbf{v}} \cdot \hat{\mathbf{p}}\right)} \label{cos}\\
\sin \phi \, \hat{\boldsymbol{\phi}} &= \frac{ v \tilde{p} + \left(\sqrt{v^2+1} -1 \right)\left(\sqrt{\tilde{p}^2+1} - 1\right)\left(\hat{\mathbf{v}} \cdot \hat{\mathbf{p}}\right)}{1 + \sqrt{v^2+1} \sqrt{\tilde{p}^2+1} + v \tilde{p} \left(\hat{\mathbf{v}} \cdot \hat{\mathbf{p}}\right)} \left( \hat{\mathbf{v}} \times \hat{\mathbf{p}}\right), \label{sin}
\end{align}
\end{widetext}
where $\tilde{p} = \left\| \mathbf{p} \right\| /m$ and $v=\left\| \mathbf{v} \right\|$. Expanding Eqs.~\eqref{cos} and \eqref{sin} in $v$ to second order around $p/m=0$ yields equations \eqref{cosphi} and \eqref{sinphi}.

\section{Coefficients in the state $\tilde{\rho}$ of Eq.~\eqref{AfterBoostTwirl}}
\label{Cs} 

The coefficients appearing in the channel in Eq. \eqref{AfterBoostTwirl} are given by
\begin{align}
c_1 &=  1 + \frac{T_2 }{4} \Bigg( \frac{1}{\kappa_v} H\left(\kappa_v\right) + \frac{1}{\kappa_p} H\left(\kappa_p\right)  \nn \\
& \qquad \qquad \qquad \qquad- \frac{3}{\kappa_v \kappa_p}  H\left(\kappa_v\right) H\left(\kappa_p\right) - 1 \Bigg), \\
c_2 &= \frac{T_1}{2} H\left(\kappa_v \right) H\left(\kappa_p \right), \\
C_1 &=  \frac{T_2}{4}  \frac{H\left(\kappa_p \right)}{\kappa_p}\Bigg( 1 - \frac{1}{\kappa_v}  H\left(\kappa_v \right)\Bigg) ,\\
C_2 &= \frac{T_2}{4} \Bigg(\frac{5}{\kappa_v \kappa_p} H\left(\kappa_v \right) H\left(\kappa_p \right) \nn \\
&   \qquad \qquad \qquad \ \ -\frac{2}{\kappa_v}H\left(\kappa_v \right)    - \frac{2}{\kappa_p}H\left(\kappa_p \right) + 1 \Bigg),\\
C_3 &= \frac{T_2}{4}  \frac{H\left(\kappa_v \right)}{\kappa_v}\Bigg( 1 - \frac{1}{\kappa_p}  H\left(\kappa_p \right)\Bigg) ,
\end{align}
where $T_n:=T^{(p)}_{n}T^{(v)}_{n}$ and
\begin{align}
T^{(v)}_{n} &:=\int_0^1 \mathrm{d}v \,  v^{2} (F(v))^n h_2 (v),  \\
T^{(p)}_{n} &:=\int_{0}^{\infty} \mathrm{d}p \, (2\pi)^{-3}(2p^0)p^2\left(\frac{p}{m}\right)^{n}f_2(p).
\end{align}
\vfill



\end{document}